# Graphical Interface for Visually Impaired People Based on Bi-stable Solenoids

Stanislav Simeonov
University “Prof. Dr. Asen Zlatarov”
Burgas, Bulgaria
e-mail: stanislav_simeonov@btu.bg

Neli Simeonova
University “Prof. Dr. Asen Zlatarov”
Burgas, Bulgaria
e-mail: neli_simeonova@btu.bg

***Abstract*— In this paper a concept for hardware realization of graphic tactile display for visually impaired peoples is presented. For realization of tactile actuators bi-stable, solenoids and PIC based control board are used. The selected algorithm for series activation of each row of display allows using minimal number of active components to set and reset the solenoids. Finally, a program algorithm of control board is discussed. The project is funded by Bulgarian National Science Fund – NSF Grant No D-ID-02/14, 2009-2013**

***Keywords-tactile display; taxel; bi-stable solenoids; visually impaired people***

## I. Introduction

Every man in the world has the need to communicate with the surrounding world. The Making computers personal and the invention of the Internet demolished the existing boundaries and started a new era in communication. Thus, every day billions of people contact with each other, share information, experience, emotions and ideas. In modern society, the access to information is fundamental right of each citizen. Unfortunately, there is a group of people for whom the use of computers and World Wide Web is bound by certain limitations. People with reduced sight perceive the surrounding world most often by direct contact and, in particular, by touching objects. Acquiring information by touch is classified into two main ways – kinetic aesthetic and tactile. The first one is usually related to the sense of location, velocity and force and it is generated by contraction of muscles and tendons. The devices designed on this basis interact with the user applying computer controlled force, thus creating an illusion for continuous and stable link to the object [1].

The perception by skin is most often related to the direct contact of the skin with the surface explored which is expressed by feeling of pressure, pulling, vibration, temperature and in some cases even pain caused by electro stimulation.

By traditional methods of presentation of information, the Braille alphabet is used. However, it is effective only when the information is in the form of text and turns out to be quite ineffective when certain graphical image is to be presented.

In recent years, we see more and more discussions on the integration of tactile displays in computer systems. With them the interpretation of graphical information will be substantially improved, thus removing the barrier between the computer and the visually impaired user.

The tactile displays are a complex electromechanic system where each pixel should be controlled individually and its activation is identified by its translation and setting at certain height. Nowadays, various methods for realization of mechanical movement of individual pixels are realized. Yobas et al. [2] used miniature electrostatic valves built on the basis of elastic thin-layer membranes of poly-silicon which expand when air pressure is applied to them and contract under electric voltage. The use of such technology reduces significantly the size of the driving mechanism. Sizes by the order of 70μm were achieved which would allow realizing tactile displays of high resolution. Nevertheless, the problems related to membrane cracking when electric voltage is applied without air pressure prevented their wide use. The diaphragm suggested by Wu et al. [3] showed significant strength and stability under pressure but the weak point of the design turned out to be the necessity to recharge the pneumatic system with compressed air and the risk of possible injures by accidental failure of the air pressure system.

Safer for the user are the systems where the driving is electric rather than pneumatic. More and more, piezoelectric linear microdrives are used for the positioning of the individual pixels [4, 7], bimetal plates [6] and alloys Shape Memory Alloy (SMA), which change their size when electric current is passed through them and restore their previous shape when current stops. [5, 8, 9, 11]. The main disadvantage of the technologies mentioned above is their high technological price which makes the prices of tactile displays based on them often higher than USD 10 000. Taking into account that most of the visually impaired people cannot afford buying such displays, it becomes necessary to search for alternative solutions. On the other hand, the development of micro-electromechanical systems (MEMS) [2] allowed crating miniature solenoids which could successfully be used in a design of a tactile display which will be much more affordable for the people with reduced sight.

The aim of the present work is to present a concept for hardware realization of graphical interface for visually impaired on the basis of micro two-positional systems.

## II. CONCEPT FOR THE DESIGN OF TACTILE GRAPHICAL DISPLAY

The use of solenoids in tactile displays is not a new idea. Some of the first prototypes were based on the same concept. [10, 12]. Despite the simplicity so far as construction is concerned, they are quite ineffective due to the large amount of energy consumed to initiate and maintain the active state of the solenoids. On the other hand, the design of graphical display with two-positional solenoids will significantly reduce the energy consumption sins it will be consumed only by turning the solenoids on and off while the plunger is detained by permanent magnets.

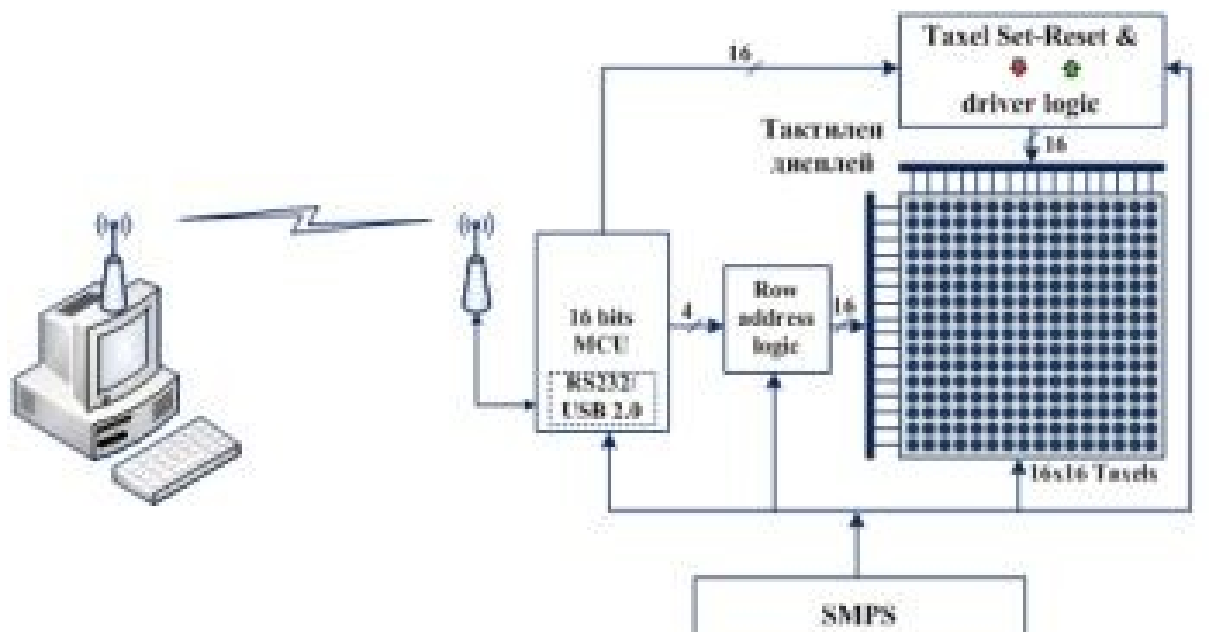

Figure 1. Block diagram of a tactile display for visually impaired people

A block diagram of a solenoid based display with resolution 16x16 taxels [7, 9] for visually impaired people is shown in Figure 1. The diagram includes the following main units:

16-bit microcontroller with integrated RS232/USB 2.0 interface;

Logical unit for addressing display rows;

Unit for initialization and resetting the taxels;

Solenoid board;

Power supply unit.

### A. Description of the block diagram

The main controlling unit of the display is a 16-bit peripheral circuit board with microcontroller PIC24FJ32GB002, product of Microchip. The microcontroller comprises hardware support for USB v2.0 standard which allows realizing the connection between the display and the user system through cable or by Wi-Fi. The choice of the 16-bit microcontroller was stipulated by the condition of simultaneous initialization and resetting the display taxels. The controller has 21 Input/Output pins which are enough for the repetitive scans of display rows simultaneously with the taxels initialization. Four of the pins at port A are used for instructions to the logical circuitry for row selection while the pins at port B are used to send signals for turning the solenoids on and off.

The row selection logic is realized by one 4-bit priority decoder 74HCT154, the outputs of which are connected to the gates of MOSFET transistors. Thus, individual rows are scanned consecutively while only 4 of the microcontroller pins are used for their selection.

The simultaneous initialization and resetting of the taxels is possible due to the principle of operation of two-positioned solenoids avoiding the excess energy consumption. One of the problems which has to be solved here is to reduce the number of active elements necessary to change voltage polarity by activation and deactivation of the solenoids. The conventional methods use half bridge circuit for each solenoid. Thus, the control of 256 taxels will require 512 transistors for the half bridge circuit or 1024 for the bridge circuit which is highly unacceptable. Due to the algorithm for the row scanning, the logic for the taxels initialization and resetting uses only 32 transistors and 16 microcontroller pins. The circuit realization is presented in Figure 2. It should be noted here that the drivers of the MOSFET transistors are not shown. Its principle of operation is as follows:

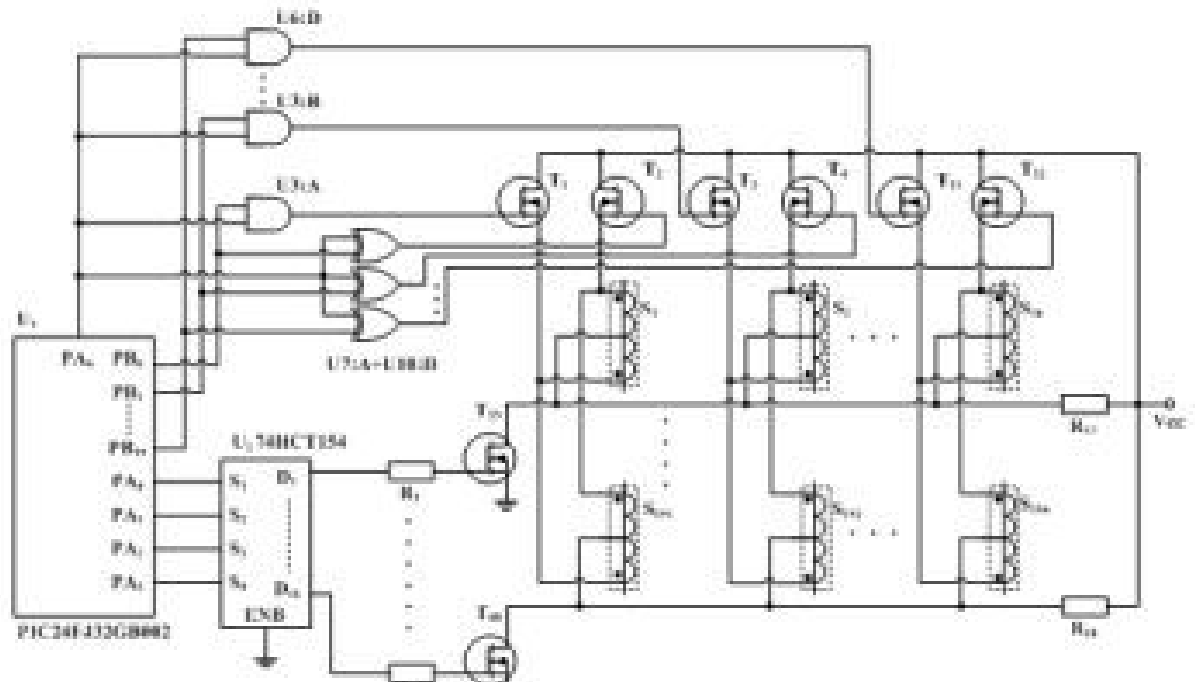

Figure 2. Operation diagram for initialization and resetting the display taxels using minimum number of active elements

### B. Initialization of the taxels

The microcontroller generates an series of pulses on pins of Port A (PA0÷PA3) which are fed to the inputs of the priority decoder 74HCT154. Due to its principle of operation, the priority decoder activates only the input corresponding to the binary combination supplied, which further sends opening signal to the transistor selecting the required row. Opening the transistor leads to a low potential fed to the solenoids so they can be now activated. Which solenoid will be activated is determined by the binary word supplied to port B – pins (PB0÷PB15) and the logical level at pin PA4. The transistors with even numbers are used to activate the solenoids and the signal to their gates is transmitted through the logical element EXCLISIVE OR (XOR). Thus, the activation occurs when logical low is supplied at pin PA4 and logical high at the corresponding pin at Port B. the transistors of uneven numbers deactivate the solenoids which is effected by supplying logical high at pin PA4 and port B, since the gates of these transistors are connected through logical elements AND. The opening of a transistor leads to electric current passing through solenoid coil and the corresponding initialization of the plunger at high position. After stopping the electric impulse, the plunger remains in this position under the influence of a permanent magnet, i.e. the there is power consumption only at the moment of its activation. By the selection of the next row, there is high potential at the common pin of the previously activated solenoids so the following opening of the transistors will not lead to activation of the non-activated taxels.

### *C. Taxels reset*

To deactivate the solenoids, a current impulse of reverse polarity should be fed and it is done by opening the transistors of uneven numbers. The resetting of the taxels is again by rows but in this case high logical level is supplied to all pins of port B. Simultaneously, logical high is supplied also to pin PA4 which allows the resetting.

Considerations on the choice of solenoids and the constructive design of the solenoid circuit board

There are several of considerations which should be taken into account for the choice of the solenoids. The first and most important one is that the solenoids should be two-positioned and their activation and deactivation must be realized through different coils. The second one is the pressing force which they can resist without mechanically resetting the plunger and the type. The third consideration is related to solenoid size, the active area and plunger design, its response time. The fourth one is the power supply – direct or pulsed current.

Presently, there are solenoids on the market with sizes by the order of millimeters. A model of the BRICON Electronics Company is shown in Figure 3, with dimensions 7x8.4x23 mm and weight of 6 g.

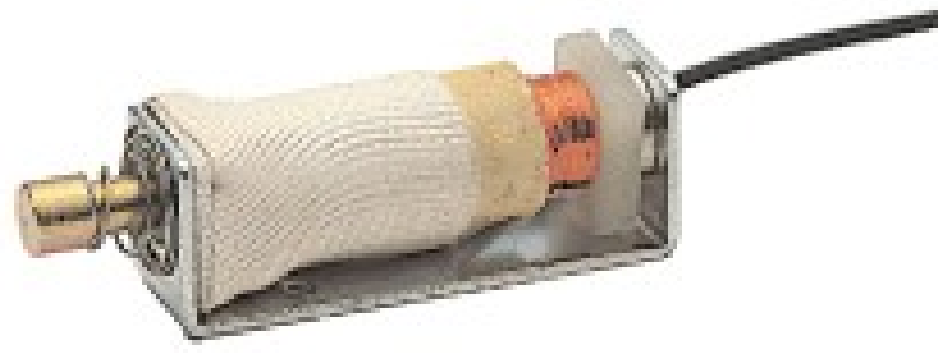

Figure 3. Two-position solenoid model SC0323L of the BRICON Company

The solenoid size is important with regard to the taxel density. The optimal utilization of space can be achieved by three-row arrangement of the solenoids and an example of this configuration is presented in Figure 4.

It can be seen that such an arrangement provides enough space for air circulation between individual solenoids, thus complying with the cooling requirements.

The pressing force which the plunger can resist to stay in active state maintained only by a permanent magnet is about 500 g, which should be enough since visually impaired people identify objects by touch but not applying big force.

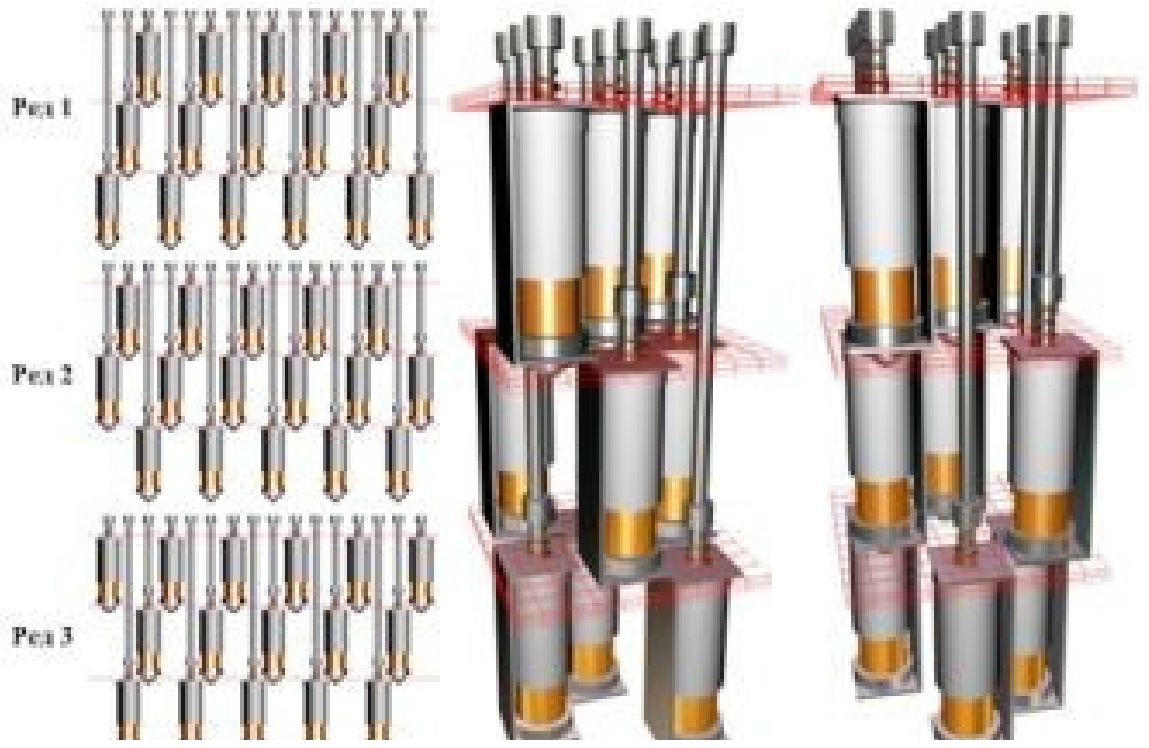

Figure 4. Arrangement of solenoids by rows and columns

The selection of control regime – either by direct current or by pulse current, is important from the point of view of providing long-term performance of the solenoid without overheating it. Usually, the catalogue data provide the value of the nominal direct current voltage for continuous performance at temperature of 20°C. If pulsed current is selected for the control, however, this voltage should be recalculated according to the coefficient δ:

$$U_P = U_{DC} \cdot \frac{1}{\delta} \qquad (1)$$

where:

UP – is the value of the pulsed voltage, V;

UDC – cThe value of the nominal constant voltage, V;

δ – coefficient of filling.

Thus, if 12 V constant voltage is necessary to deactivate the plunger, then under pulse regime the coefficient δ = 0.

## III. Specified Software

THE PIC microcontroller is of 16-bit architecture and has 32 Kb software memory and 8 Kb RAM [13]. Its speed reaches up to 16 MIPS and it is optimized for programming in C language. An example of an algorithm of microcontroller performance is presented in Figure 5. The idea of the algorithm is to make display available as soon as the power supply is switched on. On a command to bring an image to the display, the scanning of rows begins and the corresponding taxels are activated. As soon as the image is on the screen, the controller waits for the operator to give command to clear display which triggers new scanning of rows to bring them in initial state.

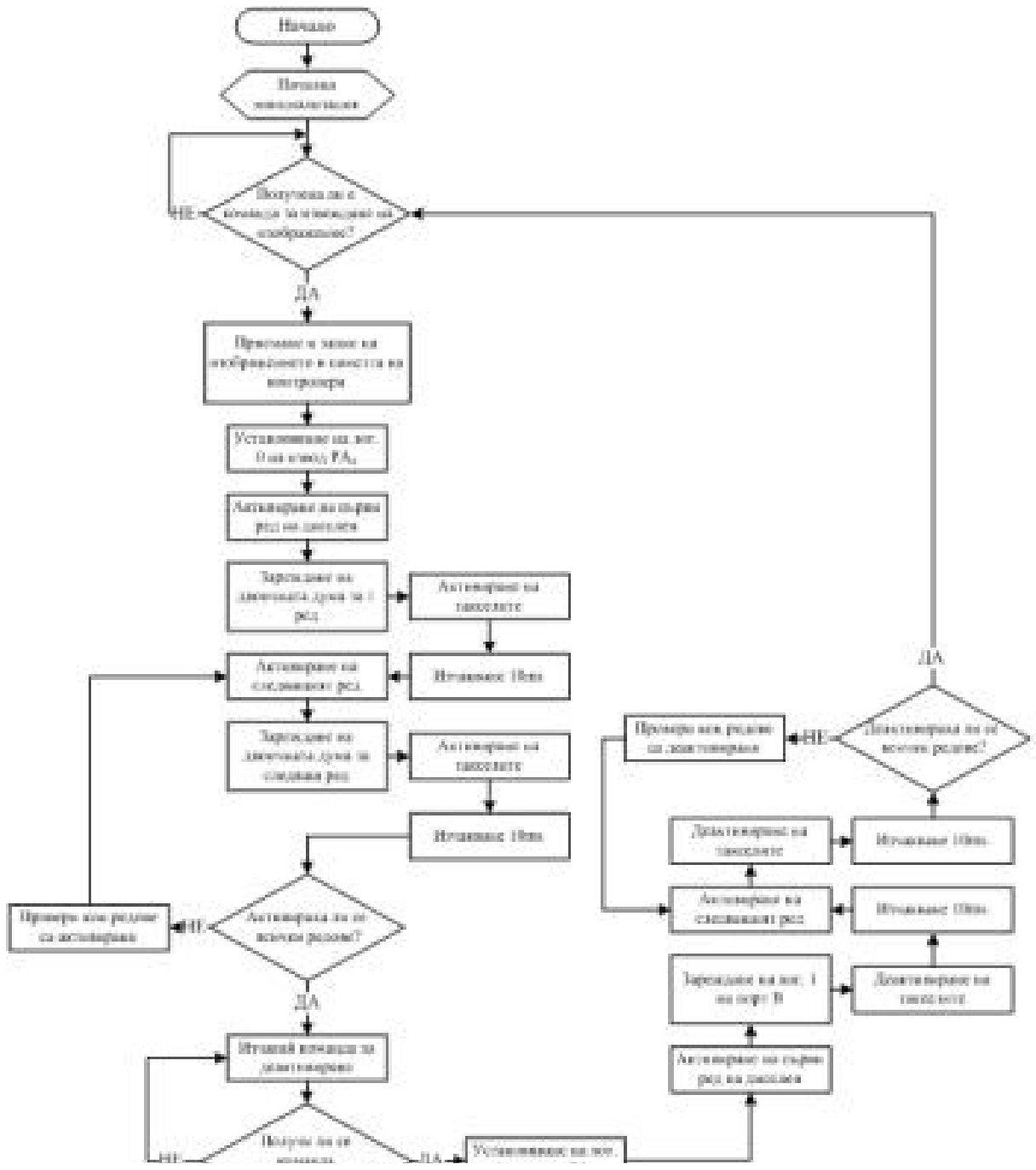

Figure 5.Program algorithm of diplay performance

## IV. CONCLUSION

In the present report, a concept for hardware realization of a graphical tactile display for visually impaired people based on two-positioned solenoids with consecutive activation and deactivation of display taxels was suggested and substantiated.

A model scheme based on 16-bit PIC microcontroller and program algorithm for effective control of activation and deactivation of solenoids was suggested using minimum hardware resources.

To optimize taxel density, solenoid arrangement in three layers was suggested to facilitate free circulation of air necessary for their cooling.

## REFERENCES

[1] G. Burdea, Force and touch feedback for virtual reality, 1996, New York, Wiley Interscience

[2] L. Yobas, D. M. Durand, G. G. Skebe, F. J. Lisy, M. A. Huff, A Novel Integrable Microvalve for Refreshable Braille Display System, Journal of microelectromechanical systems, vol. 12, no. 3, june 2003 pp. 252-263

[3] X. Wu, H. Zhu, S. Kim, M. G. Allen, A portable pneumatically-actuated refreshable braille cell, The 14th International Conference on Solid-State Sensors, Actuators and Microsystems, Lyon, France, June 10-14, 2007

[4] Hyun-Cheol Cho, Byeong-Sang Kim, Jung-Jun Park, Jae-Bok Song, Development of a Braille Display using Piezoelectric Linear Motors, SICE-ICASE International Joint Conference 2006,Oct. 18-2 1, 2006 in Bexco, Busan, Korea

[5] T. Matsunaga, K. Totsu, M. Esashi and Y. Haga, Tactile Display for 2-D and 3-D Shape Expression Using SMA Micro Actuators, Proceedings of the 31d Annual International IEEE EMBS Special Topic Conference on Microtechnologies in Medicine and Biology Kahuku, Oahu, Hawaii u 12 - 15 May 2005

[6] Gi-Hun Yang, Ki-Uk Kyung, M.A. Srinivasan, Dong-Soo Kwon, Quantitative Tactile Display Device with Pin-Array Type Tactile Feedback and Thermal Feedback, Proceedings of the 2006 IEEE International Conference on Robotics and AutomationOrlando, Florida - May 2006

[7] Genkov, D. “An Approach for Finding Proper Packet Size in IPv6 Networks”, CompSysTech’11, Vienna, 2011.

[8] Karastoyanov, D. Control of robots and other mechatronic systems. Sofia, Academy Publishing House, 2010, ISBN 987-954-322-415-9

[9] Hermes Hernandez, Enrique Preza, and Ramiro Vel´azquez, Characterization of a Piezoelectric Ultrasonic Linear Motor for Braille Displays, 2009 Electronics, Robotics and Automotive Mechanics Conference

[10] Feng Zhao, Keishi Fukuyama and Hideyuki Sawada, Compact Braille display using SMA wire array, The 18th IEEE International Symposium on Robot and Human Interactive Communication Toyama, Japan, Sept. 27-Oct. 2, 2009

[11] D. R. Chaves, I. L. Peixoto, M. F. Vieira, A.C.O. Lima, C. J. de Araújo, Microtuators of SMA for Braille display system, MeMeA 2009 - International Workshop on Medical Measurements and Applications Cetraro, Italy May 29-30, 2009

[12] Sarah F. Frisken-Gibson, Paul Bach-Y-Rita,Willis J. Tompkins, John G. Webster, A 64-Solenoid, Four-Level Fingertip Search Display For The Blind, Ieee Transactions On Biomedical Engineering, Vol. Bme-34, No. 12, December 1987

[13] A. Milev, S. Krasimirov, B. Naidenov , Security analysis in wireless sensor networks International scientific conference Proceedings UNITECH'11 Gabrovo 18-19.11.2011, vol.1 pp 316-321